\documentclass[preprint]{ptptex}
\usepackage{wrapft}

\usepackage{graphicx}

\pubinfo{Vol.~***, No.~*, *** 200*}
\preprintnumber[3cm]{KUNS-****}

\title{Gauge Theory and Dirac Operator on Noncommutative Space II}
\subtitle{Minkowskian and Euclidean Cases}

\author{Yoshinobu \textsc{Habara}}

\inst{Department of Physics, Kyoto University, Kyoto 606-8502, Japan}

\recdate{***, ** 200*}

\abst{In the preceding paper [arXiv:hep-th/0604217], we construct the Dirac operator and the integral on the canonical noncommutative space. As a matter of fact, they are ones on the noncommutative torus. In the present article, we introduce the method to extend to the Minkowskian and Euclidean cases. As a concluding remark, we present a geometrical notion of our gauge theory.}

\newcommand{\tr}{\operatorname{tr}}

\begin{document}
\maketitle

As for the canonical noncommutative space, 
\begin{align}
	& [\hat{x}^{\mu},\hat{x}^{\nu}]=i\theta^{\mu \nu}, 
	\qquad \mu =0,1,\cdots ,d-1, \nonumber \\
	& \eta^{\mu \nu}=\text{diag}(\pm 1,1,\cdots ,1), 
	\label{eqs:1}
\end{align}
since the generators $\hat{x}^{\mu}$ are represented in terms of the infinite-dimensional matrices, the ordinary trace, the most familiar map from matrices to c-number, as the volume integration is evidently divergent. Therefore we need to introduce the regularized trace which is known as the Dixmier trace.\cite{rf:connes,rf:landi} In Ref.~\citen{rf:y3}, we construct a volume integration written as the trace of matrix and a Dirac operator $D^2=\frac{1}{2\theta}\sum_{\mu =0}^{d-1}(\hat{x}^{\mu})^2$ which regulates it and plays a role as volume element. Furthermore we show that the gauge fields $A_{\mu}(\hat{x})$, the functions on the noncommutative space, are expanded in plane waves like 
\begin{align}
	A_{\mu}(\hat{x})=\sum_{k\in \boldsymbol{Z}}
	A_{\mu ,k}e^{ik_{\mu}\hat{x}^{\mu}},
	\label{eqs:2}
\end{align}
and they obey the ortho-normality condition: 
\begin{align}
	\int e^{-ik_{\mu}\hat{x}^{\mu}}\cdot e^{il_{\mu}\hat{x}^{\mu}}
	=\prod_{\mu =0}^{d-1}\delta_{k_{\mu}l_{\mu}}.
	\label{eqs:3}
\end{align}
As is obvious from (\ref{eqs:2}), this system is invariant under $\hat{x}^{\mu}\! \to \! \hat{x}^{\mu}+2\pi n^{\mu},\> n^{\mu}\in \boldsymbol{Z}$ and so we find that these gauge fields are those on the noncommutative torus.

The purpose of this letter is to formulate a theory of gauge field on the noncommutative Minkowskian/Euclidean space-time. Although it would be done in a straightforward way to extend the wave number $k_{\mu}\! \in \! \boldsymbol{Z}$ of the plane waves to real number $\boldsymbol{R}$: 
\begin{align}
	A_{\mu}(\hat{x})=\int d^d\vec{k}\> 
	A_{\mu}(\vec{k})e^{ik_{\mu}\hat{x}^{\mu}}, 
	\label{eqs:4}
\end{align}
it is not trivial they follow the ortho-normality condition: 
\begin{align}
	\int e^{-ik_{\mu}\hat{x}^{\mu}}\cdot e^{il_{\mu}\hat{x}^{\mu}}
	=\delta^d(\vec{k}-\vec{l}).
	\label{eqs:5}
\end{align}
Then  we would like to show it by defining a new volume integration $\int \! =\! \tr_{\Omega}$. For simplicity, we take $d\! =\! 2$ in the following, but it is easy to extend to the case for the general even-dimensional space-time $d=2r$.

We define a Dirac operator just like as in Ref.~\citen{rf:y3}, 
\begin{align}
	D^2=\frac{1}{\theta}\left(a_1^{\dagger}a_1+\frac{1}{2}\right), 
	\label{eqs:6}
\end{align}
whose N-th eigenvalue is $\mu_N(D^2)\! =\! \frac{1}{\theta}(N+\frac{1}{2})$ and it's degeneracy $m_N(D^2)\! =\! 1$. Also we define a Dixmier trace $\tr_{\Omega}$ and volume integration $\int$ on Minkowskian/Euclidean noncommutative space as 
\begin{align}
	& \tr_{\Omega}a\equiv \lim_{N\to \infty} \sum_{n=0}^{N}
	m_n(a)\mu_n(a), \nonumber \\
	& \int a\equiv \frac{1}{\theta}\tr_{\Omega}D^{-2}a.
	\label{eqs:7}
\end{align}
In general, for the case of $d$-dimensional space-time, $\int a\! \equiv \! \frac{\theta^{d/2}}{d/2}\tr_{\Omega}D^{-d}a$. This integral diverges for the unit matrix $1$ in contrast to the case of the torus\cite{rf:y3}: 
\begin{align}
	\int 1=\frac{1}{\theta}\tr_{\Omega}D^{-2}
	=\frac{1}{\theta}\lim_{N\to \infty}\sum_{n=0}^N1\times 
	\theta \left(n+\frac{1}{2}\right)^{-1}=\lim_{N\to \infty}\log N.
	\label{eqs:8}
\end{align}
However we will find that this divergence turns out to be convergence in the sense of the hyperfunctions, i.e. Dirac delta function, when we discuss the ortho-normality of the plane waves.

Functions on the noncommutative space are expanded as 
\begin{align}
	a(\hat{x})=\int d^2\vec{k}\> a(k_0,k_1)e^{ik_0\hat{x}^0+ik_1\hat{x}^1}.
	\label{eqs:9}
\end{align}
To show (\ref{eqs:5}), we evaluate the left hand side of it, and find that the result almost coincides with that of Ref.~\citen{rf:y3} in it's form: 
\begin{align}
	\int & e^{-ik_0\hat{x}^0-ik_1\hat{x}^1}\cdot 
	e^{il_0\hat{x}^0+il_1\hat{x}^1} \nonumber \\
	& =\lim_{N\to \infty}\sum_{n=0}^N\frac{1}{n+\frac{1}{2}}
	\langle n|e^{-ik_0\hat{x}^0-ik_1\hat{x}^1}\cdot 
	e^{il_0\hat{x}^0+il_1\hat{x}^1}|n\rangle \nonumber \\
	& =e^{-\frac{\theta}{4}(k_0-l_0)^2-\frac{\theta}{4}(k_1-l_1)^2
	+i(k_0l_1-k_1l_0)} \nonumber \\
	& \qquad \qquad \times \lim_{N\to \infty}\sum_{n=0}^{N}
	\frac{1}{n+\frac{1}{2}}
	F\left([-n],[1],\textstyle \frac{\theta}{2}(k_0-l_0)^2
	+\frac{\theta}{2}(k_1-l_1)^2\right), 
	\label{eqs:10}
\end{align}
except for the wave number $k_{\mu}$ being real number $\boldsymbol{R}$. Here, $F([-n],[1],z)$ is the Laguerre polynomial,
\begin{align*}
	F([-n],[1],z)\equiv \sum_{m=0}^n\frac{(-1)^mz^m}{m!}\frac{n!}{m!(n-m)!}
	=L_n(z).
\end{align*}
As is well known, $L_n(z)e^{-\frac{1}{2}z}$ organize the ortho-normal basis in $0\! \leq \! z\! <\! +\infty$: $\int_{0}^{+\infty}dz\> e^{-z}L_n(z)L_m(z)=\delta_{nm}$. Also (\ref{eqs:10}) includes the very form of the basis, $\delta_N(z)\! \equiv \! \sum_{n=0}^N\frac{1}{n+\frac{1}{2}}e^{-\frac{1}{2}z}L_n(z),\> z\! \geq \! 0$, excluding the phase factor $e^{i(k_0l_1-k_1l_0)}$ which is irrelevant to the behavior in the limit $N\! \to \! \infty$. Therefore the sequence $\delta_N(z)$ can be normalized for any $N$, i.e. we can regard it as the sequence which satisfiy $\int_{0}^{+\infty}dz\> \delta_N(z)=1$. And, in view of (\ref{eqs:8}), we have $\lim_{N\to \infty}\delta_N(z\! =\! 0)\! =\! +\infty$.
\begin{figure}[htb]
	\centerline{\includegraphics*[width=10cm,angle=0]{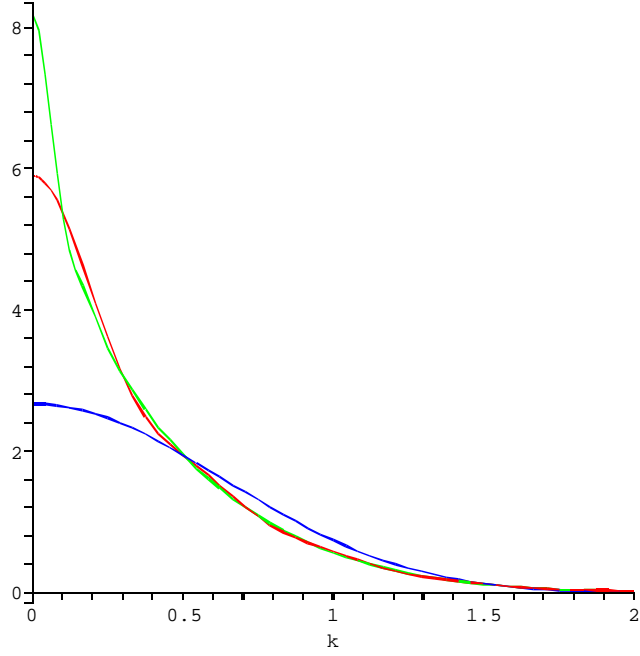}}
	\caption{The behavior of $\delta_N(z)\! =\! \sum_{n=0}^N
	\frac{1}{n+\frac{1}{2}}e^{-\frac{1}{2}z}L_n(z)$. 
	The blue, red and green lines denote 
	the cases for $N=1,50$ and $500$ respectively.}
	\label{fig:1}
\end{figure}
As is shown in Fig.~\ref{fig:1}, the larger $N$ becomes, the more $\delta_N(z)$ sharpens at $z\! =\! 0$ and the more convergent it becomes at $z\! \to \! +\infty$. Next, let us investigate the behavior of $\delta_N(z)$ at $N\! \to \! \infty$ by utilizing the genetating function of $L_n(z)$, 
\begin{align}
	\sum_{n=0}^{\infty}L_n(z)t^n=\frac{e^{-\frac{t}{1-t}z}}{1-t}, 
	\qquad |t|<1.
	\label{eqs:11}
\end{align}
Integrating the both sides of (\ref{eqs:11}) with $t$, we obtain 
\begin{align}
	\sum_{n=0}^{\infty}\frac{1}{n+1}L_n(z)t^n
	=\int^t dt^{\prime}\> 
	\frac{e^{-\frac{t^{\prime}}{1-t^{\prime}}z}}{1-t^{\prime}}
	=-e^z\text{Ei}\left(-\frac{z}{1-t}\right).
	\label{eqs:12}
\end{align}
Here, we have $\frac{1}{n+1}$ in front of $L_n(z)$ which differs from $\frac{1}{n+\frac{1}{2}}$ of $\delta_N(z)$, which is irrelevant when $N\! \to \! \infty$. $\text{Ei}(-x)\! =\! -\int_{x}^{+\infty}dq\> \frac{e^{-q}}{q}$ is the well-known exponential integral. Then we find $\lim_{N\to \infty}\delta_N(z)=0, z>0$ if we take $t\! \to \!1_-$ with $z$ keeping fixed. From these considerations, we acquire $\lim_{N\to \infty}\delta_N(z)=\delta(z)$, i.e. the right hand side of $(\ref{eqs:10})$ converges in the sense of Dirac delta function, and (\ref{eqs:5}) is fulfilled. As a result, the integral $\int$ picks out the zero mode: 
\begin{align}
	\int a(\hat{x})=a(k_{\mu}=0).
	\label{eqs:13}
\end{align}
Of course the unitary invariance can be checked in the exactly same way as in Ref.~\citen{rf:y3}. And the action $S_{\text{YM}}$ of the gauge field on the $4$-dimensional Minkowskian/Euclidean noncommutative space is found to be 
\begin{align}
	S_{\text{YM}} 
	& =\frac{1}{4}\int [A_{\mu}(\hat{x}),A_{\nu}(\hat{x})]
	[A^{\mu}(\hat{x}),A^{\nu}(\hat{x})]
	+\frac{1}{4}\int [A_{\mu}(\hat{x}^{\prime}),A_{\nu}(\hat{x}^{\prime})]
	[A^{\mu}(\hat{x}^{\prime}),A^{\nu}(\hat{x}^{\prime})] 
	\nonumber \\
	& \qquad \qquad -\frac{1}{2}\theta_{\mu \nu}^{-1}\theta^{-1 \mu \nu}
	\int 1, \nonumber \\
	& =-2\int d^4\vec{k}d^4\vec{l}d^4\vec{m}d^4\vec{n}\> 
	A_{\mu}(\vec{k})A_{\nu}(\vec{l})A^{\mu}(\vec{m})A^{\nu}(\vec{n})
	\sin \frac{k_{\mu}l_{\nu}\theta^{\mu \nu}}{2}
	\sin \frac{m_{\mu}n_{\nu}\theta^{\mu \nu}}{2} \nonumber \\
	& \qquad \qquad \times 
	\cos \frac{(k+l)_{\mu}(m+n)_{\nu}\theta^{\mu \nu}}{2} 
	\delta^4(\vec{k}+\vec{l}+\vec{m}+\vec{n}) 
	-\frac{1}{2}\theta_{\mu \nu}^{-1}\theta^{-1 \mu \nu}\delta^4(0). 
	\label{eqs:14}
\end{align}
It is obvious that $\int 1\! =\! \lim_{N\! \to \! \infty}\log N\! =\! \delta^4(0)$.

Finally, let us make some comments. Our gauge theory (\ref{eqs:14}) lacks a kinetic term and also ``momentum" field $\Pi_{\mu}\! \sim \! \partial_tA_{\mu}$ conjugate to $A_{\mu}$. Furthermore, as we guess also from Refs.~\citen{rf:y1,rf:y2,rf:y3}, the algebra of our noncommutative space is, in principle, the Heisenberg algebra, to be exact, the linear transformation of it. Therefore, we may consider the second quantization method should be 
\begin{align}
	[A_{\mu}(\vec{k}),A_{\nu}(\vec{l})]=i\theta^{-1}_{\mu \nu}
	\delta^4(\vec{k}-\vec{l}), 
	\label{eqs:15}
\end{align}
in the operator formalism, just in the same way as the space-time itself. Then, if we take a proper linear combinations of $A_{\mu}(\vec{k})$ like $\hat{x}^{\mu}$, they turn out to be the creation and annihilation operators of the harmonic oscillator system.

Next, for the gauge theory on the ordinary commutative space, we have the term as $A_{\mu}^a(\vec{k})A_{\nu}^b(\vec{l})A^{\mu c}(\vec{m})A^{\nu d}(\vec{n})f^{abe}f^{cde}$ instead of (\ref{eqs:14}) which includes the structure constant $f^{abc}$ of the Lie algebra $[T^a,T^b]=if^{abc}T^c$, while there exist in (\ref{eqs:14}) the noncommutative structure of the space-time and a mysterious ``cosmological" constant term $-\frac{1}{2}\theta_{\mu \nu}^{-2}\delta^4(0)$. This constant term comes from the first quantization (\ref{eqs:1}), i.e. the space-time noncommutativity itself, and is proportional to the space-time ``volume". In order to delete it, it is sufficient to naively redefine the field strength as $\hat{\mathcal{F}}_{\mu \nu}\! \equiv \! [\nabla_{\mu},\nabla_{\nu}]+i\theta^{-1}_{\mu \nu}=-[A_{\mu}(\hat{x}),A_{\nu}(\hat{x})]$. Here let us think of its meaning. As we can see from the construction of the supersymmetric gauge theory in Ref.~\citen{rf:y2}, the genuine component of the vector multiplet is $v_{\mu}(\hat{x})$ in $A_{\mu}(\hat{x})=\hat{p}_{\mu}+v_{\mu}(\hat{x})$ by nature. And, as is well known, $A_{\mu}(\hat{x})=\hat{p}_{\mu}\! =\! -\theta^{-1}_{\mu \nu}\hat{x}^{\nu}$ is a solution of the equation of motion $[A_{\nu}[A^{\mu},A^{\nu}]]=0$ of the action $S_{\text{YM}}$. If we regard gauge fields as a fluctuation around the classical solution, the field strength is written as 
\begin{align}
	\hat{F}_{\mu \nu} & =[\nabla_{\mu},\nabla_{\nu}]
	=-[A_{\mu}(\hat{x}),A_{\nu}(\hat{x})]-i\theta^{-1}_{\mu \nu} 
	\nonumber \\
	& =-[\hat{p}_{\mu},v_{\nu}(\hat{x})]+[\hat{p}_{\nu},v_{\mu}(\hat{x})]
	-[v_{\mu}(\hat{x}),v_{\nu}(\hat{x})], 
	\label{eqs:16}
\end{align}
which looks very like that on the ordinary commutative space and lacks the term $\theta^{-1}_{\mu \nu}$. Moreover, introducing a fluctuation $X^{\mu}(\hat{x})$ of the space-time coordinate itself as $\hat{p}_{\mu}+v_{\mu}(\hat{x})\! \equiv \! -\theta^{-1}_{\mu \nu}(\hat{x}^{\nu}+X^{\nu}(\hat{x}))$, we have 
\begin{align}
	& \hat{y}^{\mu}=\hat{x}^{\mu}+X^{\mu}(\hat{x}), \nonumber \\
	& [\hat{y}^{\mu},\hat{y}^{\nu}]=i\theta^{\mu \nu}
	+[\hat{x}^{\mu},X^{\nu}(\hat{x})]+[X^{\mu}(\hat{x}),\hat{x}^{\nu}]
	+O(X^2),
	\label{eqs:17}
\end{align}
and if we regard the above gauge field as a gauge parameter, $X^{\mu}(\hat{x})$ seems to be a parameter of the infinitesimal general coordinate transformation and, so that the algebraic structure (\ref{eqs:1}) may invariant under this transformation at the first order in $X^{\mu}$, from (\ref{eqs:17}), 
\begin{align}
	[\hat{x}^{\mu},X^{\nu}(\hat{x})]+[X^{\mu}(\hat{x}),\hat{x}^{\nu}]=0 
	\quad \Longleftrightarrow \quad 
	[\hat{p}_{\mu},v_{\nu}(\hat{x})]-[\hat{p}_{\nu},v_{\mu}(\hat{x})]=0, 
	\label{eqs:18}
\end{align}
must be satisfied. Namely $\hat{F}_{\mu \nu}\! =\! -[v_{\mu}(\hat{x}),v_{\nu}(\hat{x})]$ from (\ref{eqs:16}), which is the same form as the naively redefined $\mathcal{F}_{\mu \nu}\! =\! -[A_{\mu}(\hat{x}),A_{\nu}(\hat{x})]$. This assumption might seem to be plausible if we imagine that our flat noncommutative space would be a ``target" space on a curved noncommutative manifold. Our gauge theory is {\it not} invariant under (\ref{eqs:17}) since we consider the ``transformation" (\ref{eqs:17}) on the flat space-time, and the gauge parameter looks like the gauge field. Of course, it is invariant under the unitary transformation which can be thought as a subclass of ``transformation" (\ref{eqs:17}) keeping (\ref{eqs:1}) invariant. Therefore, we can conclude that ignoring the ``cosmological constant" term which is proportional to the volume of the noncommutative space is equivalent to keeping the algebraic structure of the noncommutative space under the infinitesimal general coordinate transformation. From these considerations, if we preserve the invariance of the structure of the flat Minkowskian/Euclidean noncommutative space (\ref{eqs:1}) under the diffeomorphism and consider that the action (\ref{eqs:14}) describes the second order quantum fluctuation to the flat noncommutative geometry, we can eliminate the ``cosmological constant" term $-\frac{1}{2}\theta_{\mu \nu}^{-2}\int 1$. Conversely saying, we would think the gauge theory (\ref{eqs:14}) is nothing but describing the weak geometric fluctuation to the flat noncommutative space. When we carry out the quantization via the path integral formalism, we may need to restrict the integration region as to $A_{\mu}$ within the subclass which satisfies (\ref{eqs:18}).


\end{document}